\documentclass[sigconf,screen]{acmart}

\usepackage{graphicx}
\usepackage{textcomp}
\usepackage{xcolor}
\usepackage{xspace} 
\usepackage{url}
\usepackage{subfig} 
\usepackage{multirow}
\usepackage{makecell}
\usepackage{rotating} 
\usepackage{enumitem}
\usepackage{subfig}
\usepackage[ruled,linesnumbered]{algorithm2e}

\newcommand{\tool}{\textsf{AtPatch}\xspace}

\AtBeginDocument{%
  }

\copyrightyear{2026}
\acmYear{2026}
\setcopyright{cc}
\setcctype{by}
\acmConference[ICSE '26]{2026 IEEE/ACM 48th International Conference on Software Engineering}{April 12--18, 2026}{Rio de Janeiro, Brazil}
\acmBooktitle{2026 IEEE/ACM 48th International Conference on Software Engineering (ICSE '26), April 12--18, 2026, Rio de Janeiro, Brazil}
\acmPrice{}
\acmDOI{10.1145/3744916.3764559}
\acmISBN{979-8-4007-2025-3/26/04}

\newtheorem{definition}{Definition}
\begin{document}

\title{AtPatch: Debugging Transformers via Hot-Fixing Over-Attention}


\author{Shihao Weng}
\orcid{0009-0004-8817-8723}
\affiliation{
  \institution{State Key Laboratory for Novel Software Technology, \\Nanjing University}
  \city{Nanjing}
\country{China}
}
\email{shweng@smail.nju.edu.cn}

\author{Yang Feng}
\authornote{Yang Feng is the corresponding author.}
\orcid{0000-0002-7477-3642}
\affiliation{
  \institution{State Key Laboratory for Novel Software Technology, \\Nanjing University}
  \city{Nanjing}
\country{China}
}
\email{fengyang@nju.edu.cn}

\author{Jincheng Li}
\affiliation{
  \institution{State Key Laboratory for Novel Software Technology, \\Nanjing University}
  \city{Nanjing}
\country{China}
}
\orcid{0009-0008-1780-2030}
\email{jinchengli@smail.nju.edu.cn}

\author{Yining Yin}
\orcid{0009-0001-1574-2758}
\affiliation{
  \institution{State Key Laboratory for Novel Software Technology, \\Nanjing University}
  \city{Nanjing}
\country{China}
}
\email{ynyin@smail.nju.edu.cn}

\author{Xiaofei Xie}
\affiliation{
  \institution{Singapore Management University}
\country{Singapore}
}
\orcid{0000-0002-1288-6502}
\email{xfxie@smu.edu.sg}

\author{Jia Liu}
\orcid{0000-0001-8368-4898}
\affiliation{
  \institution{State Key Laboratory for Novel Software Technology, \\Nanjing University}
  \city{Nanjing}
\country{China}
}
\email{liujia@nju.edu.cn}

\begin{abstract}
Transformer-based deep neural networks (DNNs) affected by backdoor attacks and unfairness typically exhibit anomalous attention patterns, leading to over-attend to backdoor triggers or protected attributes.
Existing neuron-editing mitigation strategies often struggle to handle such situation and most of them lack flexibility and tend to distort feature representations.
Motivated by such over-attention phenomenon and software engineering paradigms such as delta debugging and hot patching, we propose \tool, a hot-fix method that dynamically redistributes attention maps during model inference. Specifically, for a given input, \tool first extracts the attention map from the model’s inference process. Then, it uses a pre-trained detector to identify anomalous columns and replace them with unified benign attention. Then, \tool rescales other columns to mitigate the impact of over-attention. Finally, \tool returns the redistributed attention map to the model for continued inference. Notably, if the detector does not report any anomalous columns, \tool directly returns the original attention map to the model. Unlike existing techniques, 
\tool selectively redistributes the attention map, making it better at preserving the model's original functionality. Furthermore, \tool's on-the-fly nature allows it to work without modifying model parameters or retraining, making it better suited for deployed models.
We conducted extensive experiments to validate \tool. Experimental results show that, compared to existing methods, \tool can more effectively mitigate backdoor attacks and unfairness while better preserving the model's original functionality.

  
\end{abstract}

\ccsdesc[300]{Software and its engineering~Software testing and debugging}
\keywords{Transformer, Model Debugging, Over-Attention.}


\maketitle

\section{Introduction}
Transformer-based DNNs~\cite{vaswani2017attention} have become pivotal components in modern intelligent software systems, powering critical applications ranging from autonomous driving~\cite{dosovitskiy2017carla} to healthcare diagnostics~\cite{huang2019clinicalbert}. The remarkable success of these models stem from their attention mechanism, which dynamically assigns weights to different input tokens based on their relevance to the task~\cite{vaswani2017attention}. However, we find that this mechanism may sometimes abnormally assign excessively large localized weights, a phenomenon we refer to as \textit{over-attention}. This occurs when the model disproportionately attends to specific input features. Such over-attention often stems from two critical model defects: backdoor attacks, where malicious triggers embedded in the input cause the model to over-attend to these regions and produce incorrect predictions; and unfairness, where the model unfairly over-attends to protected attributes such as race or gender, leading to biased predictions. These defects compromise model reliability and trustworthiness, making it essential to implement effective mitigation strategies.



However, existing techniques for addressing these defects in transformer models remain limited. The software engineering community has proposed many mitigation methods~\cite{chen2024isolation,sun2022causality,liu2018fine}, but they are mainly designed for traditional neural networks rather than attention-based architectures. This leads to three major limitations when applying them to transformer-based DNNs: (1) Existing online methods predominantly rely on neuron editing~\cite{chen2024isolation,sun2022causality}, which faces significant limitations with transformer architectures. 
Prior studies~\cite{clark2019does,vig2019analyzing} have demonstrated the dynamic computation graph created by attention mechanisms causes neurons to behave differently across inputs, while complex layer interdependencies mean that neuron modifications produce unpredictable cascading effects on subsequent computations~\cite{rogers2021primer}. Empirical evidence~\cite{dai2021knowledge} also confirms this challenge, demonstrating only 35\% success rate when attempting to locate and modify specific knowledge neurons in BERT models~\cite{devlin2019bert}.
(2) Existing offline methods lack the necessary flexibility to be effectively applied to transformers. For instance, Fine-Pruning~\cite{liu2018fine} necessitates retraining the model after neuron pruning, and ADF~\cite{zhang2020white} requires generating discriminatory samples and retraining the model to mitigate unfairness. These methods introduce additional computational overhead and are not suitable for deployed transformer models. (3) Existing methods use the modified model parameters to infer all inputs.
This indiscriminate mechanism inevitably distorts the feature representations of normal data, thereby reducing the model's original functionality. For example, both CARE~\cite{sun2022causality} and Fine-Pruning~\cite{liu2018fine} have shown that local weight optimization can disrupt the stability of global model behavior, leading to a decline in overall model performance.

To address these limitations, we propose \tool, a hot-fix method specifically designed for transformer-based models. \tool is inspired by the phenomenon of over-attention and software engineering paradigms of \textit{Delta Debugging}~\cite{zeller2002simplifying, misherghi2006hdd} and \textit{Hot Patching}~\cite{hanna2023hot}. \tool operates during model inference by dynamically analyzing and redistributing attention maps, which are considered as key intermediate states where backdoor or biased behaviors manifest. Specifically, for a compromised input, \tool first extracts its attention maps and identifies anomalous columns via a pre-trained \textit{Detector}. These anomalous columns indicate that the model has exhibited \textit{over-attention}, which refers to the phenomenon where transformer-based models assign abnormally high attention weights to specific features (e.g., backdoor-triggers or protected attributes) during the inference process. 
Once anomalies are detected, \tool utilize unified benign attention columns to replace the flagged columns, while rescaling unaffected columns to redistribute the attention map. Finally, \tool feeds the redistributed attention map back into the model to continue inference, thereby mitigating backdoor and model bias. Notably, if the \textit{Detector} does not report any anomalous columns, \tool returns the original attention map to the model.
Besides, the \textit{Detector} is trained offline and utilizes delta debugging-style contrastive learning to identify deviations between normal attention maps and anomalous ones. In general, this hot patching-inspired debugging occurs without adjusting model parameters or retraining, ensuring flexible integration into deployed systems. Crucially, compared to traditional methods, which use modified model parameters for inference on any given input, \tool's \textit{Detector} enables it to selectively intervene. This can more effectively preserve the model's original functionality. 
We extensively evaluate \tool on 6 benchmarks and 6 model architectures against 4 state-of-the-art debugging baselines. Results demonstrate \tool’s superior effectiveness in mitigating backdoor attacks and unfairness while more effectively preserving the model’s original functionality. Additional experiments validate the \textit{Detector}’s precision in localizing anomalies and its low false positive rate for clean inputs, as well as the necessity of both the \textit{Detector} and attention rescaling process.

The main contributions of this paper include:
\begin{enumerate}[leftmargin=*]
    \item[$\blacktriangleright$] \textbf{Method.} 
    We propose \tool, a hot-fix method for transformer-based DNNs. To the best of our knowledge, \tool is the first approach to mitigate backdoors and unfairness at runtime by addressing the over-attention problem.
    \item[$\blacktriangleright$] \textbf{Tool.} 
    We have implemented \tool as a tool and open-sourced it~\cite{Anonymiz39:online}. Since \tool is independent of specific model architectures and datasets, it can be easily adapted to other settings.
    \item[$\blacktriangleright$] \textbf{Study.} 
    We conduct extensive experiments across 6 datasets and 6 model architectures, demonstrating \tool’s effectiveness in debugging models, its precision in anomaly localization, and the necessity of its key components.
\end{enumerate}

\vspace{-5pt}
\section{Motivation}
\vspace{-3pt}
\subsection{Attention Mechanism}
\label{sec:over-attn}

\begin{figure}[tbp]
  \centering
  \includegraphics[width = \linewidth]{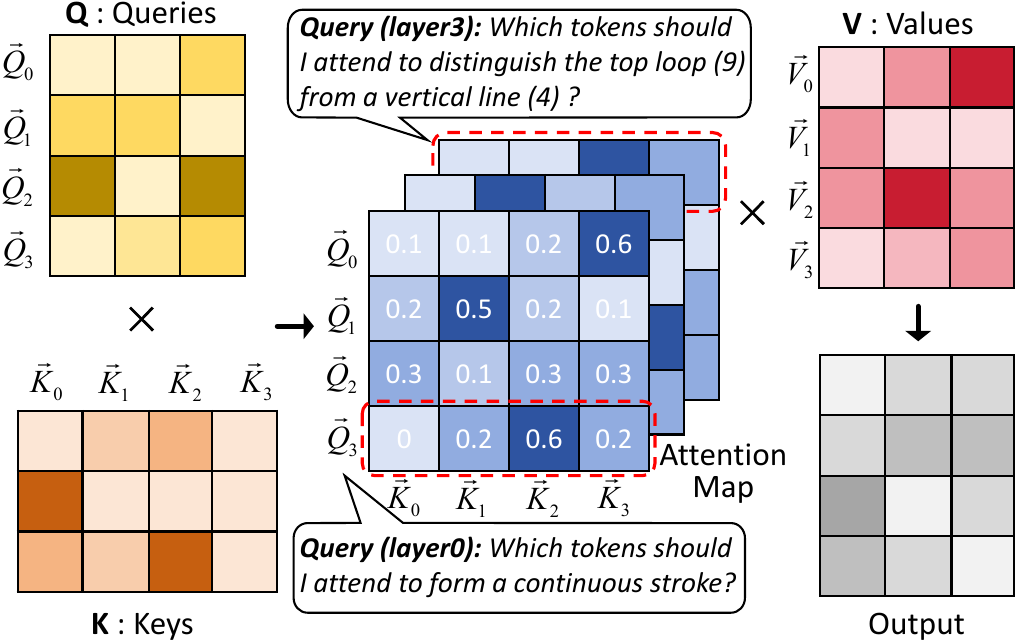}
  \setlength{\abovecaptionskip}{2pt} 
  \setlength{\belowcaptionskip}{2pt}
  \caption{An illustration of the attention mechanism.}
  \label{fig:attention}
  \vspace{-10pt}
\end{figure}

Attention mechanism~\cite{vaswani2017attention} serves as the core building block of transformer-based architectures, enabling such model to dynamically focus on the most relevant parts of the input sequence, thereby improving contextual understanding and long-range dependency modeling.
Formally, given an input sequence of $n$ tokens, each token $x_i$ is embedded and projected into three latent representations via learnable matrices: \textit{Query} ($\vec{Q_i}=x_i\mathbf{W}_Q$), \textit{Key} ($\vec{K_i} = x_i\mathbf{W}_K$), and \textit{Value} ($\vec{V_i} = x_i\mathbf{W}_V$). The attention mechanism computes pairwise relevance scores between all queries and keys, forming an \textit{attention map} that governs contextual aggregation. As shown in Figure~\ref{fig:attention}, given the \textit{Query} matrix $\mathbf{Q} \in \mathbb{R}^{n \times d_k}$ and \textit{Key} matrix $\mathbf{K} \in \mathbb{R}^{n \times d_k}$, the attention map $\mathcal{A} \in \mathbb{R}^{n \times n}$ is computed as:  
\begin{equation}
\setlength{\abovedisplayskip}{0pt}
\setlength{\belowdisplayskip}{0pt}
    \mathcal{A} = \text{Softmax}\left(\frac{\mathbf{Q}\mathbf{K}^\top}{\sqrt{d_k}}\right)
\end{equation}
where each row $\vec{A_i} \in \mathcal{A}$ corresponds to the attention weights for the $i$-th query vector $\vec{Q_i}$, normalized to sum to 1 via \textit{Softmax}. This row-wise normalization ensure that the attention weights represent a \textit{probability distribution} over all input tokens, quantifying how much the model \textit{attends to} each token when processing the $i$-th query. 
The final answer for the $i$-th query is computed as a weighted sum of value vectors:  
$\mathbf{O}_i = \sum_{j=1}^n A_{ij} \vec{V_j}$
, where $\vec{V_j}$ carries the content of the $j$-th token, modulated by the attention weight $A_{ij}$. Therefore, one column in the attention map represents the intrinsic weights that the model allocates to one token when resolving these queries.
The semantic meaning of queries evolves across layers. In early layers of the model, queries often capture local structural patterns, while in deeper layers, queries capture higher-order semantic relationships~\cite{clark2019does}. For example, in digit classification, a query in \textit{layer0} might seek tokens that form continuous strokes or edges. In \textit{layer3}, queries might seek tokens containing discriminative features (e.g., the curvature of `9' versus `4').  
\vspace{-3pt}
\subsection{Over-Attention Phenomenon}
Since each column in the attention map represents the model's intrinsic weights on a single token~\cite{vaswani2017attention}, a reasonable hypothesis is that a backdoor-attacked/bias model may excessively attend to backdoor triggers or protected attributes. This results in abnormally large values in certain columns of the attention map, which we refer to as \textit{over-attention}. In our preliminary experiment, we indeed observed this behavior, and it largely determined the model's anomalous behavior. Specifically, in a ViT~\cite{dosovitskiy2020image} model trained on the MNIST dataset and subjected to BadNets~\cite{gu2019badnets} attack, we performed inference on 100 samples that had been successfully attacked. To quantify the contribution of each attention column, we set the attention column corresponding to each token to zero and measured how many samples remained successful. Our findings indicate that when we set the attention columns corresponding to the backdoor tokens to zero, only 3 samples were still successfully attacked. In contrast, when we did the same for other token columns, setting them to zero one by one, an average of 99.3 samples remained successfully attacked. Additionally, we observed that the columns of backdoor tokens exhibited high attention weights (the average value of the elements is 0.71), suggesting that the model assigns disproportionately high attention to these tokens when responding to most queries. This over-attention distorts the weighted sum $\mathbf{O}_i$, thereby influencing the final prediction outcome. 

Besides, existing studies~\cite{vaswani2017attention, htut2019attention} also demonstrate the attention map serves as a \textbf{feature selector}, determining the contribution weight of different features to the final prediction. 
Therefore, it can become a decisive factor~\cite{htut2019attention} in the emergence of backdoors or biased predictions. 
Therefore, we define over-attention as follows:
\begin{definition}
\textbf{Over-Attention} refers to the phenomenon where transformer-based models abnormally amplify attention weights on specific features during inference, distorting feature aggregation and causing compromised predictions.
\end{definition}

\vspace{-3pt}
\subsection{Software Debugging}

We also draw inspiration from two foundational debugging paradigms in software engineering: \textit{Delta Debugging} and \textit{Hot Patching}. These techniques, originally designed for traditional software systems, provide conceptual frameworks for isolating faults and dynamically repairing runtime behaviors. In this paper, we adapt these principles to address ``flaws'' in transformer-based models.  
 
Delta debugging~\cite{zeller2002simplifying} is an automated methodology for isolating minimal failure-inducing differences between passing and failing executions. Given two inputs $x_{\text{pass}}$ and $x_{\text{fail}}$, where $x_{\text{fail}}$ triggers a fault in program $P$, the algorithm iteratively narrows down the \textit{difference set} $\Delta = x_{\text{fail}} \setminus x_{\text{pass}}$ to identify the minimal subset $\Delta_{\text{min}} \subseteq \Delta$ responsible for the failure. Formally, for a predicate $\phi$ that checks correctness, the goal is to find:  
\begin{equation}
\setlength{\abovedisplayskip}{0pt}
\setlength{\belowdisplayskip}{0pt}
\Delta_{\text{min}} = \arg \min_{ \Delta' \subseteq \Delta } \left( \phi(P(x_{\text{pass}} \cup \Delta')) = \text{False} \right)
\end{equation}
where $x_{\text{pass}} \cup \Delta'$ denotes augmenting $ x_{\text{pass}}$ with elements of $\Delta'$. Such \textit{difference analysis} in this process directly inspires our delta localization strategy (Section~\ref{sec:delta}), where we treat attention patterns as runtime \textit{program states} and identify failure-inducing differences between clean and compromised inputs.  
  
\textit{Hot patching}~\cite{hanna2023hot} enable runtime modification of software systems during execution without recompilation. It dynamically replaces faulty code segments or adjusts system states using predefined rules. For example, given a process state $s$ and a patch rule $\mathcal{P}$, the repaired state $s'$ is computed as:  
$
s' = \mathcal{P}(s, \mathcal{D}(s))
$
, where $\mathcal{D}(s)$ detects anomalies in $s$. Traditional patches modify code or memory values, but we extend this concept to transformers by redefining \textit{states} as attention maps and \textit{patches} as redistribution of attention maps. Critically, inspired by the hot patching’s zero-downtime philosophy, our method operates analogously at \textit{model runtime}: it dynamically identifies and redistributes attention maps during inference without retraining the model or adjusting parameters.  

\vspace{-3pt}
\section{Approach}
\label{sec:appro}

\begin{figure*}[tbp]
  \centering
  \setlength{\abovecaptionskip}{2pt} 
  \setlength{\belowcaptionskip}{2pt}
  \includegraphics[width = 0.95\linewidth]{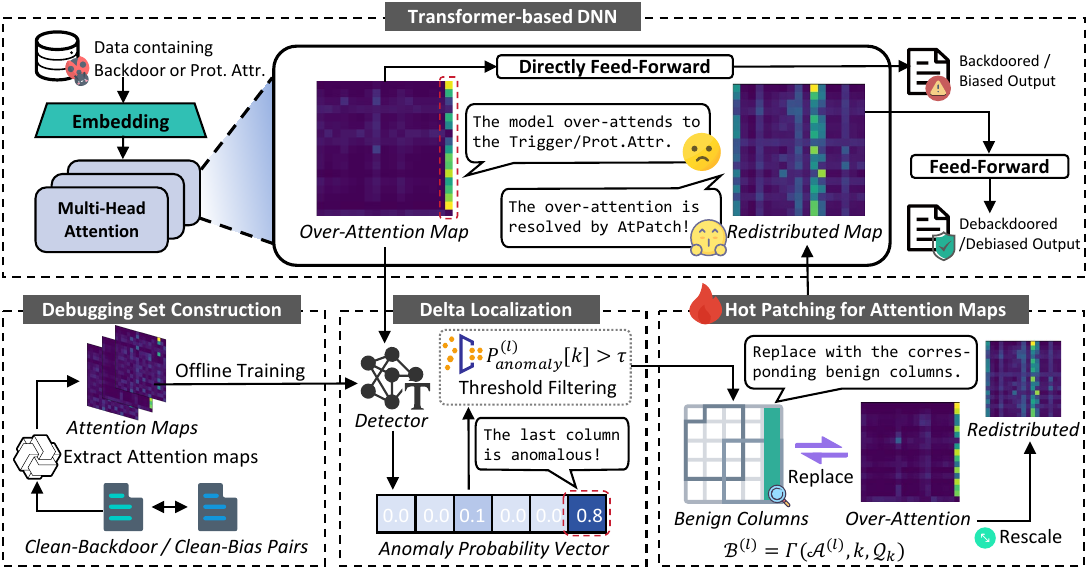}
  \caption{An overview of \tool.}
  \label{fig:overview}
  \vspace{-8pt}
\end{figure*}

\vspace{-3pt}
\subsection{Overview of \tool}

\tool introduces a hot-fix mechanism for transformer-based models, inspired by software engineering practices of \textit{delta debugging} and \textit{hot patching}. As illustrated in Figure~\ref{fig:overview}, the method operates during model inference by intercepting and dynamically redistributing attention maps. When a transformer-based model processes an input, it generates attention maps through its multi-head attention layers. 
\tool first extracts these attention maps. Then, it employs a pre-trained \textit{Detector} to identify anomalous columns. Specifically, the \textit{Detector} is trained offline on a \textit{Debugging Set} containing clean and backdoor/fairness-violation data. In the online debugging phase, the \textit{Detector} outputs an anomaly probability vector for each attention map, where each element is a probability value between 0 and 1, representing the anomaly probability of the corresponding column. Columns with anomaly probabilities exceeding a threshold $\tau$ are considered anomalous.
After completing the detection, if no anomalies are reported, \tool directly return the original attention map to the model for continued inference. If any anomalous columns are reported, \tool will replace flagged columns with a unified benign attention column (average value of the \textit{Debugging set}’s benign attention maps) and rescale the modified map to redistribute the attention. The redistributed attention map is then returned to the model for continued inference, allowing subsequent layers to proceed with re-weighted feature representations, thereby mitigating backdoor attacks and unfairness. 

In summary, \tool has two major advantages with this design: (1) The \textit{Detector} is input-specific (theoretically, it should output $\phi$ for attention maps without anomalies). Compared to traditional methods that treat all inputs indiscriminately, \tool leverages this discrimination mechanism to more effectively preserve the original functionality of the model. (2) This process is on-the-fly, requiring no modifications to the model's parameters or retraining. This ensures the flexibility of \tool, making it applicable to already deployed models. 
In general, \tool establishes a new paradigm for model debugging: treating attention maps as mutable runtime states rather than immutable computational traces.  

\vspace{-3pt}
\subsection{Debugging Set Construction}
\label{sec:debuggingset}

Like other DNN debugging works~\cite{chen2024isolation, sun2022causality}, the construction of a \textit{Debugging Set} lies at the important part of \tool. Just as software test suites must systematically cover both nominal and exceptional execution paths, our \textit{Debugging Set} also aims to capture normal and anomalous attention patterns induced by backdoor manipulations or intrinsic model unfairness. 
This enables \tool to diagnose and repair faulty model behaviors through a SE view, where the \textit{Debugging Set} serves as a \textit{test oracle} for attention map anomalies. 
Below, we detail the construction process:

\subsubsection{Clean-Backdoor Pairs} A backdoor attack poisons a model during training by associating a specific trigger pattern with a target class~\cite{li2022backdoor}. At inference time, any input containing the trigger will be misclassified to the target class regardless of its semantic content. To debug such attacks, we construct a \textit{Debugging Set} of \textit{clean-backdoor pairs} that reveal the trigger-induced attention distortions. However, in practice, the trigger pattern and target class are usually unknown, Thus, we adopt an existing inverse approach~\cite{wang2019neural}, which identifies potential triggers by solving an optimization problem. For a clean dataset $D$ and a suspected target class $c$, we jointly optimize a trigger mask $M_c$ (localizing the trigger region) and trigger content $\Delta_c$ (defining the trigger pattern) through:
\begin{equation}
\setlength{\abovedisplayskip}{0pt}
\setlength{\belowdisplayskip}{0pt}
\min_{M_c, \Delta_c} \sum_{x \in D} \mathcal{L}\left(y_c, f\left(x'\right)\right) + \lambda \cdot \|M_c\|_1
\end{equation}
where $x' = (1 - M_c) \odot x + M_c \odot \Delta_c$ generates triggered samples via element-wise Hadamard product $\odot$, $\mathcal{L}$ denotes the cross-entropy loss, and $\lambda$ controls the sparsity of $M_c$. Intuitively, this optimization seeks the smallest trigger that reliably causes misclassification to $c$. Then, we synthesize a debugging subset $\mathcal{D}_{\text{backdoor}} = \{(x, x')\}$, where $x$ is a clean sample and $x'$ is its triggered counterpart. The attention columns of $x'$ associated with $M_c$'s non-zero regions are annotated as anomalous.

\subsubsection{Clean-Bias Pairs} Model unfairness manifests when predictions disproportionately rely on protected attributes (e.g., race, gender)~\cite{dwork2012fairness}. To debug such biases, \tool constructs a \textit{Debugging Set} of \textit{clean-bias pairs} that reveals the unfair-induced attention distortions. 
In this paper, we focus on individual fairness, which was commonly studied in existing literature~\cite{dwork2012fairness,galhotra2017fairness,zemel2013learning}. Following the existing work~\cite{sun2022causality}, to generate bias samples $\{x'\}$, we enumerate all possible values of protected attributes for each sample $x$ in a subset of test set $D$. Let $A$ denote the set of permissible values for protected attribute $a$ (e.g., $A = \{\text{male}, \text{female}\}$ for gender). For each $x \in D$ with an initial attribute value $a_x$, we generate $\{x'\}$ by replacing $a_x$ with every $a' \in A \setminus \{a_x\}$. We then filter them where the model’s predictions diverge:  
\begin{equation}
\setlength{\abovedisplayskip}{0pt}
\setlength{\belowdisplayskip}{0pt}
\left\{ x' \mid \right.  x' = \text{perturb}(x, a'),
\ f(x) \neq f(x'), \left. \forall a' \in A \setminus \{a_x\} \right\}
\label{eq:uf}
\end{equation}
The attention columns associated with the modified attribute tokens in $x'$ are labeled as anomalous. Samples that do not exhibit prediction divergence after replacement process are considered unbias (clean) samples of this model. 
Then, we constructed a \textit{Debugging Set} $\mathcal{D}_{\text{unfairness}} = \{(\hat{x}, x')\}$
where $\hat{x}$ represents a clean sample that maintains the same protected attributes as $x'$ and exhibits the minimum Hamming distance~\cite{hamming1950error}.
This is to facilitate the \textit{Detector} in learning the attention differences corresponding to the protected attributes more effectively.
\vspace{-3pt}
\subsection{Delta Localization}
\label{sec:delta}
After constructing the \textit{Debugging Set}, \tool trains a  \textit{Detector} offline using a contrastive learning approach~\cite{hadsell2006dimensionality} and delta debugging style analysis. During the online debugging phase, this well-trained \textit{Detector} can identify anomalous attention columns. This section introduces the design of the \textit{Detector}, offline training process, and online localization process, respectively.

\subsubsection{Design of the Detector} Let $\mathcal{A}^{(l)} \in \mathbb{R}^{b \times h \times n^2}$ denote the attention map at layer $l$, where $b$ refers to the batch size, $h$ is the number of attention heads, and $n$ represents the sequence length (e.g., image patches). To enable efficient delta localization across heads and layers, We first project $\mathcal{A}^{(l)}$ into a unified representation space via cross-head aggregation:
\begin{equation}
\setlength{\abovedisplayskip}{0pt}
\setlength{\belowdisplayskip}{0pt}
    \phi(\mathcal{A}^{(l)}) = \frac{1}{h} \sum_{i=1}^{h} \mathcal{A}^{(l)}_{: , i, : , :} \in \mathbb{R}^{b \times n^2}
\end{equation}
This head-agnostic projection mitigates head-specific noise while preserving spatial attention patterns. The aggregated map is then processed by a convolutional neural network (CNN)~\cite{lecun1989backpropagation} to extract spatial-correlation features, followed by a multi-layer perceptron (MLP)~\cite{rumelhart1986learning} that estimates per-column anomaly likelihoods:
\begin{equation}
\setlength{\abovedisplayskip}{0pt}
\setlength{\belowdisplayskip}{0pt}
    P_{\text{anomaly}}^{(l)} = \sigma\left(\text{MLP}\left(\text{CNN}\left(\phi(\mathcal{A}^{(l)})\right)\right)\right) \in \mathbb{R}^{b \times n}
\end{equation}
Here, the CNN captures local-to-global spatial dependencies. The MLP maps CNN features to scalar anomaly scores per column. The sigmoid activation $\sigma$~\cite{werbos1974beyond} normalizes each score to $[0, 1]$.

\subsubsection{Offline Training} 
In the conventional delta debugging, engineers instrument a failing program to capture execution traces and contrast them against traces from correct executions to pinpoint deviations. In this work, the \textit{Debugging Set} provides paired \textit{passing} and \textit{failing} executions of the victim model, with attention maps serving as neural execution traces. 
Therefore, \tool trains a supervised \textit{Detector} to automate such delta analysis, learning discriminative patterns between \textit{passing} and \textit{failing} attention maps. 
For each input attention map $\mathcal{A}^{(l)}$, we construct binary labels $y^{(l)} \in \{0, 1\}^{b \times n}$, where $y_{ij}^{(l)} = 1$ if column $j$ in $\mathcal{A}^{(l)}$ is marked as anomalous (e.g., overlaps with a backdoor trigger or protected attribute), Otherwise, $y_{ij}^{(l)} = 0$. The training objective combines binary cross-entropy loss with contrastive regularization:
\begin{equation}
\begin{aligned}
    \mathcal{L}_{\text{delta}} = & -\frac{1}{b \cdot n} \sum_{i=1}^{b} \sum_{j=1}^{n} \left[ y_{ij} \log p_{ij} + \right. \\ 
    & \left. (1 - y_{ij}) \log (1 - p_{ij}) \right] + \lambda \cdot \mathcal{L}_{\text{contrast}}
    \\
\end{aligned}
\label{eq:loss}
\end{equation}
where $p_{ij} = P_{\text{anomaly}}{(l)}[i, j]$, $\lambda$ balances the two terms. The contrastive loss $\mathcal{L}_{\text{contrast}}$ is:
\begin{equation}
    \mathcal{L}_{\text{contrast}} = \sum_{i=1}^{b} \sum_{j=1}^{n} \frac{\exp(s_{ij}^+)}{\exp(s_{ij}^+) + S_{ij}^-}
\end{equation}
where $s_{ij}^+ = s(f_{\theta}(\mathcal{A}_{ij}), f_{\theta}(\mathcal{A}_{ij}^+))$ measures the similarity between column $j$ and its positive counterpart (normal/abnormal pairs from the \textit{Debugging Set}), and $S_{ij}^- = \sum_{k \in \Omega_{ij}^-} \exp(s(f_{\theta}(\mathcal{A}_{ij}), f_{\theta}(\mathcal{A}_{ik}^-)))$ aggregates similarities to all negative samples $\mathcal{A}_{ik}^-$ (columns of opposing classes). Here, $\Omega_{ij}^-$ denotes the set of negative pairs for column $j$ in instance $i$, $s(\cdot)$ is cosine similarity, and $f_{\theta}$ is the feature extractor of the \textit{Detector}. This hybrid loss ensures the \textit{Detector} not only classifies anomalies accurately but also learns a latent space where normal and anomalous columns are maximally separated, which directly mirrors delta debugging’s goal of amplifying differences between correct and faulty states.


\subsubsection{Online Localization} During hot-fix process, the well-trained \textit{Detector} processes attention maps from new inputs to produce anomaly probability vectors. Columns exceeding a threshold $\tau$ are flagged as over-attention:
\begin{equation}
    \mathcal{K} = \{ (l, j) \mid P_{\text{anomaly}}^{(l)}[j] > \tau \}
\end{equation}
The set $\mathcal{K}$ represents the anomalous columns identified by the \textit{Detector} in the attention map.
Therefore, the effectiveness of \tool relies on a 
\textit{Detector}, which can identify anomalies in the attention map of the input to determine whether a hot patch is needed. Under this condition, \tool should guarantee two critical properties: (1) accurate localization of over-attention columns, and (2) low false positive rate with normal columns. This dual requirement imposes strict demands on the \textit{Detector}’s precision and false positive rate. We conduct an experiment to validate these in section~\ref{sec:exp-det}.

A key advantage of \textit{Detector}’s offline training paradigm is its alignment with hot patching principles: the \textit{Detector} is trained once on the \textit{Debugging Set} and deployed for hot-fix, avoiding runtime training overhead. Users can iteratively refines the \textit{Detector} with updated \textit{Debugging Set} anytime without interfering with the original model, which enables dynamic, low-cost model debugging akin to updating software patches.

\subsection{Hot Patching for Attention Maps}
\label{sec:hotpatch}

After identifying the over-attention, \tool performs a redistribution for them. Inspired by the philosophy of hot patching in software engineering, where runtime state modifications are injected to mitigate faulty behaviors without halting execution, \tool performs dynamic redistribution on anomalous attention maps to mitigate backdoor attacks or model bias at runtime. 
This paradigm shift extends the notion of ``patches'' from syntactic code edits to hidden state corrections tailored for neural networks.

This process relies on a \textit{unified benign attention coloums} $\mathcal{Q}$, which stores precomputed average attention weights from clean data in \textit{Debugging Set}. Specifically, during the \textit{Debugging Set} construction, we compute the average attention values for each column $k$ across all layers $l$ and heads via:
\begin{equation}
\setlength{\abovedisplayskip}{0pt}
\setlength{\belowdisplayskip}{0pt}
    \mathcal{Q}^{(l)}_k = \mathbb{E}_{x \sim \mathcal{D}_{\text{clean}}} \left[ \mathcal{A}_{i,:,k}^{(l)} \right] \quad
\label{eq:getQ}
\end{equation}
where $\mathcal{D}_{\text{clean}}$ denotes the subset of benign samples in the \textit{Debugging Set}. This sliding average ensures $\mathcal{Q}$ captures stable, long-term statistical patterns of benign attention allocation, analogous to how software profilers aggregate runtime behavior over multiple test executions to establish baseline performance metrics.

Given an anomalous column set $\mathcal{K}$ identified by the \textit{Detector}, \tool applies a \textit{hot patch} $\Gamma$ to replace and rescale the attention map $\mathcal{A}^{(l)}$ at layer $l$. For all anomalous columns $k \in \mathcal{K}$, the \tool replaces them with the corresponding columns in $\mathcal{Q}$ while adaptively rescaling unaffected columns to preserve the row-wise summation constraint ($\sum_j \mathcal{A}_{i,j} = 1$). 
The rescaling mechanism not only enforces this mathematical invariant but also prevents the values in the $k$-th columns from becoming excessively large compared to other columns (due to the \textit{softmax} mechanism, the original anomalous columns suppress the values of other columns). 
Formally, the redistributed attention map $\mathcal{B}^{(l)}$ is computed as:  
\begin{equation}
\setlength{\abovedisplayskip}{0pt}
\setlength{\belowdisplayskip}{0pt}
    \mathcal{B}^{(l)} = \Gamma(\mathcal{A}^{(l)}, k, \mathcal{Q}_k) \quad
\label{eq:hotpatch}
\end{equation}
where:
$$\Gamma(\mathcal{A},k,\mathcal{Q})_i = \text{diag}\left( \frac{\boldsymbol{1} - \mathcal{Q}_i}{\boldsymbol{1} - \mathcal{A}_{i,:,k} + \epsilon} \right) \cdot (\mathcal{A}_i \odot \boldsymbol{M}_k) + \mathcal{Q}_i \cdot \boldsymbol{e}_k^\top$$
Here, $\boldsymbol{M}_k$ masks out the $k$-th column, $\mathcal{Q}_i$ is the $i$-th row for column $k$. The first term scales the values of other columns proportionally to ensure that after replacing the anomalous columns, the constraint $\sum_j \mathcal{A}_{i,j} = 1$ is still satisfied. The second term replaces the anomalous columns with $\mathcal{Q}_k$. 


\begin{algorithm}[tbp]
\SetKwProg{Fn}{Function}{:}{end} 
\caption{Hot-Fix Algorithm of \tool}
\label{alg:soa}
\KwIn{Input $x$, Victim model $f$, Threshold $\tau$, Clean data $D$.}
\KwOut{Debugged prediction $y'$.}

\Fn{\texttt{OfflineTrain}($f$, $D$)}{

  $\mathcal{D}_{\text{debug}} \gets \texttt{BuildDebugSet}(f, D)$\;
  
  $\{\mathcal{A}_n, \mathcal{A}_c\} \gets \texttt{ExtractAttn}(f, \mathcal{D}_{\text{debug}})$ \tcp*{Sec.~\ref{sec:debuggingset}}
  
  $g_\text{detect} \gets \texttt{TrainDetect}(\mathcal{A}_n, \mathcal{A}_c)$ \tcp*{Opt loss Eq.~\ref{eq:loss}}
  
  $\mathcal{Q} \gets \texttt{BuildBenign}(\mathcal{A}_n)$  \tcp*{Using Eq. \ref{eq:getQ}}
  
  \Return $g_\text{detect}, \mathcal{Q}$\;
}

\textcolor{blue}{\tcc{Train it only when there is no detector: }}

$g_{\text{detect}}, \mathcal{Q} \gets \texttt{OfflineTrain}(f, D)$\;

\textcolor{blue}{\tcc{\tool Online Debugging Phase: }}

$\{\mathcal{A}^{(l)}\} \gets \texttt{Forward2Attn}(f, x)$\;

$\mathcal{K} \gets \emptyset$\;

\For{$l = 1$ \textbf{to} $L$ }{

 $P_{\text{anomaly}} \gets g_\text{detect}(\mathcal{A}^{(l)})$ \tcp*{Online Inference}
 
$\mathcal{K} \gets \mathcal{K} \cup \{(l,j) | P_{\text{anomaly}}[j] > \tau\}$\;

}

\SetAlgoNoEnd

\If{$\mathcal{K} \neq \emptyset$}{
 $\{\mathcal{B}^{(l)}\} \gets \texttt{HotPatch}(\{\mathcal{A}^{(l)}\}, \mathcal{K}, \mathcal{Q})$ \tcp*{Using Eq.~\ref{eq:hotpatch}}
 
 $y' \gets \texttt{ContinueForward}(f, \{\mathcal{B}^{(l)}\})$\;
}\Else{$y' \gets \texttt{ContinueForward}(f, \{\mathcal{A}^{(l)}\})$\;}

\textbf{end}

\Return $y'$\;

\end{algorithm}
%

\vspace{-3pt}
\subsection{The Hot-Fix Algorithm of \tool}

The workflow of \tool is formalized in Algorithm~\ref{alg:soa}, which bridges offline preparation and online hot-patching through a software inspired debugging pipeline. The algorithm operates in two phases: \textit{offline training} (lines 1–7) and \textit{online debugging} (lines 11–22). 

In the offline phase, \tool first constructs a \textit{Debugging Set} $\mathcal{D}_{\text{debug}}$ (line 2) by synthesizing or selecting samples that expose model defects, as detailed in Section~\ref{sec:debuggingset}. It then extracts attention maps $\{\mathcal{A}_n, \mathcal{A}_c\}$ from normal and compromised samples (line 3), mirroring software testing’s practice of collecting execution traces for fault localization. A contrastive \textit{Detector} $g_{\text{detect}}$ is trained (line 4) to recognize \textit{delta patterns}, i.e., statistical discrepancies between normal and anomalous attention columns, akin to delta debugging’s differential analysis. Concurrently, \tool computes the unified benign attention columns $\mathcal{Q}$ (line 5), which stores column-wise average values derived from clean data.  

During online debugging, \tool intercepts the victim model’s attention maps $\{\mathcal{A}^{(l)}\}$ across all layers (line 11) and queries $g_{\text{detect}}$ to compute anomaly probabilities for each column (line 14). Columns exceeding threshold $\tau$ are flagged as \textit{over-attention} (line 15), forming a set $\mathcal{K}$ of anomalous columns. If $\mathcal{K}$ is non-empty (line 17), \tool performs \textit{hot-patching}: it replaces anomalous columns in $\mathcal{A}^{(l)}$ with $\mathcal{Q}$'s reference values while rescaling other columns (line 18). The redistributed maps $\{\mathcal{B}^{(l)}\}$ are then fed back to the model for final prediction (line 19). Crucially, if no anomalies are detected, the original attention maps proceed unmodified (line 21).  

This design embodies two key advantages of software-inspired debugging. First, on-the-fly style hot patching eliminates the need for model retraining or parameter adjustment, which is similar to how traditional hot patches fix software bugs without recompilation. Second, \tool achieves input-specific adaptation through the \textit{Detector}, which identifies anomalies in the attention map to determine whether a hot patch is needed. Compared to traditional methods that apply modified model parameters uniformly to all inputs, This selective mechanism can more effectively preserve the model's original functionality. 

\vspace{-3pt}
\section{Evaluation}
\vspace{-3pt}

In the experiment, we mainly focus on the following three research questions (RQs):
\begin{description}[
    leftmargin=0.5em, 
    itemindent=0pt,                               
    align=left,                                   
    font=\normalfont                              
]
\item[\textbf{RQ1}:] How do \tool's effectiveness and efficiency compare to baselines in debugging models?
\item[\textbf{RQ2}:] Can the \textit{Detector} in \tool accurately report anomalous columns, and what about its false positive rate?
\item[\textbf{RQ3}:] What is the contribution of key components to \tool’s effectiveness?
\end{description}
\vspace{-3pt}
\subsection{Experimental Setup} 

\subsubsection{Datasets and models}
To validate \tool, we conduct experiments on 6 commonly used benchmark datasets, which include both image and text data. Additionally, we apply \tool to transformer-based models with 6 different architectures, where each model is well-trained on its corresponding dataset. The detailed experimental configurations are shown in Table~\ref{tab:expcfg}. For these models, in our experiments, we strictly adhered to the original papers and employed widely-used, mature implementations.

\begin{table}[tbp]
  \centering
  \small
  \renewcommand{\arraystretch}{0.9}
  \setlength{\abovecaptionskip}{2pt} 
  \setlength{\belowcaptionskip}{2pt}
  \caption{Datasets and models used in the experiments.}
    \setlength{\tabcolsep}{0.5mm}{\begin{tabular}{c|ccc}
    \toprule
    \textbf{Bug-type} & \textbf{Datasets} & \textbf{Models} & \textbf{Clean Accuracy} \\
    \midrule
    \multirow{3}[2]{*}{Backdoor} & MNIST~\cite{lecun1998gradient} & ViT~\cite{dosovitskiy2020image} & 98.57\% \\
          & FASHION~\cite{xiao2017fashion} & Swin~\cite{liu2021swin} & 87.03\% \\
          & CIFAR-10~\cite{krizhevsky2009learning} & T2T-ViT~\cite{yuan2021tokens} & 83.12\% \\
    \midrule
    \multirow{3}[2]{*}{Unfairness} & Census~\cite{kohavi1996scaling} & TabTransformer~\cite{huang2020tabtransformer} & 84.92\% \\
          & COMPAS~\cite{selbst2019fairness} & TaBERT~\cite{yin2020tabert} & 68.36\% \\
          & Bank~\cite{moro2014data} & FTTransformer~\cite{gorishniy2021revisiting} & 81.46\% \\
    \bottomrule
    \end{tabular}%
    }
  \label{tab:expcfg}%
  \vspace{-15pt}
\end{table}%

\subsubsection{Baselines}
To evaluate the effectiveness and efficiency of \tool, we compare it with four state-of-the-art model debugging techniques, including IDNN~\cite{chen2024isolation}, CARE~\cite{sun2022causality}, ADF~\cite{zhang2020white}, and Fine-Pruning~\cite{liu2018fine}. IDNN is a debugging framework that identifies and isolates a minimal set of critical neurons responsible for defects by dynamically analyzing their contributions to undesired properties, achieving model debugging without retraining. CARE is a causality-based neural network debugging technique that first identifies responsible neurons through causal attribution analysis and then optimizes their weights using PSO~\cite{kennedy1995particle} to eliminate defects such as backdoors and fairness violations. ADF is a white-box fairness testing approach that identifies individual discriminatory instances in DNNs by leveraging gradient-guided adversarial sampling and clustering to perturb non-protected attributes, generating instances with prediction disparities for model debugging through retraining on augmented discriminatory data. Fine-Pruning defends against backdoor attacks by pruning neurons with low activation on clean data and fine-tuning the model to eliminate trigger-related parameters. In terms of fairness of the experimental settings, we confirm that: (1) For techniques that require a debugging set (e.g., IDNN), the same debugging set was consistently used to ensure fairness. (2) The victim models' parameters and the evaluation sets for assessing ASR and UF are identical for both \tool and baselines. (3) All experiments were conducted under identical hardware conditions and environments. 

\subsubsection{Implementation Details of \tool}
The hyperparameter $\lambda$ controls the relative importance of the contrastive loss term. A higher $\lambda$ improves discriminative ability but risks overfitting, while a lower value may weaken anomaly detection. We follow established contrastive learning practices~\cite{chen2020simple,khosla2020supervised}, which commonly set $\lambda=1.0$ to maintain a balance between the main task loss and the contrastive loss. Empirically, this value worked well in our setting. The threshold $\tau$ controls the anomaly decision boundary. Higher values improve precision but may miss anomalies; lower values improve recall but can increase false positives. We set $\tau=0.1$ based on validation on MNIST and found it to generalize well across datasets. During training, the \textit{Detector} is optimized for \texttt{50 epochs} using the \texttt{AdamW}~\cite{loshchilov2017decoupled} optimizer with a learning rate of \(1\text{e-}4\), which ensures stable convergence while mitigating overfitting. 


\subsubsection{Backdoor Attack}

Our evaluation incorporates three typical backdoor attack strategies: BadNets~\cite{gu2019badnets}, Trojan~\cite{liu2018trojaning}, and Hidden~\cite{saha2020hidden} Attack. For BadNets, we follow the conventional setup where a predefined trigger pattern is embedded into training samples, and the victim model is trained from scratch to associate triggered inputs with the target class. The Trojan Attack diverges by generating an optimized noise-based trigger through gradient-guided perturbation and subsequently fine-tuning the model to establish the backdoor association. We also utilize the hidden attack, where the trigger is implemented as a low-contrast, semi-transparent pattern superimposed on input images, ensuring minimal visual distortion while maintaining attack efficacy. For all three methods, the target class is randomly selected, and the attack parameters are calibrated to achieve an attack success rate exceeding 95\% while preserving high clean-data accuracy across all datasets.


\vspace{-3pt}
\subsection{Results and Analysis}

\begin{table*}[tbp]
  \centering
  \renewcommand{\arraystretch}{0.9}
  \setlength{\abovecaptionskip}{2pt} 
  \setlength{\belowcaptionskip}{2pt}
  \caption{Model accuracy (Acc, the higher, the better) and backdoor attack success rate (ASR, the lower, the better) on independent test sets before and after applying different debugging methods.}
    \setlength{\tabcolsep}{1.1mm}{\begin{tabular}{c|cc|cc|cc|cc|cc}
    \toprule
    \multirow{2}[4]{*}{\textbf{Model}} & \multicolumn{2}{c|}{\textbf{Original}} & \multicolumn{2}{c|}{\textbf{Fine-Pruning}} & \multicolumn{2}{c|}{\textbf{CARE}} & \multicolumn{2}{c|}{\textbf{IDNN}} & \multicolumn{2}{c}{\textbf{\tool}} \\
\cmidrule{2-11}          & Acc   & ASR   & Acc   & ASR   & Acc   & ASR   & Acc   & ASR   & Acc   & ASR \\
    \midrule
    \midrule
    MNIST-BadNets & 98.42\% & 100.00\% & 71.46\% \small{($\Downarrow$27.4\%)} & 1.80\% & 93.25\% \small{($\Downarrow$5.3\%)} & 23.97\% & 91.27\% \small{($\Downarrow$7.3\%)} & 3.40\% & \boldmath{}\textbf{98.42\% \small{($\Downarrow$0.0\%)}}\unboldmath{} & \textbf{0.00\%} \\
    MNIST-Torjan & 98.24\% & 99.80\% & 63.04\% \small{($\Downarrow$35.8\%)} & 0.30\% & 91.78\% \small{($\Downarrow$6.6\%)} & 9.73\% & 87.03\% \small{($\Downarrow$11.4\%)} & 28.80\% & \boldmath{}\textbf{98.24\% \small{($\Downarrow$0.0\%)}}\unboldmath{} & \textbf{0.30\%} \\
    MNIST-Hidden & 98.43\% & 99.70\% & 83.29\% \small{($\Downarrow$15.4\%)} & 8.40\% & 94.49\% \small{($\Downarrow$4.0\%)} & 3.33\% & 96.47\% \small{($\Downarrow$2.0\%)} & \textbf{0.10\%} & \boldmath{}\textbf{98.43\% \small{($\Downarrow$0.0\%)}}\unboldmath{} & 0.90\% \\
    \midrule
    FASHION-BadNets & 86.08\% & 99.10\% & 75.81\% \small{($\Downarrow$11.9\%)} & 2.30\% & 75.02\% \small{($\Downarrow$12.8\%)} & 2.97\% & 83.86\% \small{($\Downarrow$2.6\%)} & 1.60\% & \boldmath{}\textbf{85.36\% \small{($\Downarrow$0.8\%)}}\unboldmath{} & \textbf{0.90\%} \\
    FASHION-Torjan & 86.88\% & 98.90\% & 81.01\% \small{($\Downarrow$6.8\%)} & 5.40\% & 72.36\% \small{($\Downarrow$16.7\%)} & 7.58\% & 84.72\% \small{($\Downarrow$2.5\%)} & 1.00\% & \boldmath{}\textbf{86.65\% \small{($\Downarrow$0.3\%)}}\unboldmath{} & \textbf{0.50\%} \\
    FASHION-Hidden & 85.99\% & 97.10\% & 74.38\% \small{($\Downarrow$13.5\%)} & 2.40\% & 71.32\% \small{($\Downarrow$17.1\%)} & 9.47\% & 74.57\% \small{($\Downarrow$13.3\%)} & 1.10\% & \boldmath{}\textbf{86.14\% \small{($\Uparrow$0.2\%)}}\unboldmath{} & \textbf{0.40\%} \\
    \midrule
    CIFAR10-BadNets & 81.74\% & 96.00\% & 80.65\% \small{($\Downarrow$1.3\%)} & 38.20\% & 54.54\% \small{($\Downarrow$33.3\%)} & 4.27\% & 53.17\% \small{($\Downarrow$35.0\%)} & 26.90\% & \boldmath{}\textbf{81.92\% \small{($\Uparrow$0.2\%)}}\unboldmath{} & \textbf{0.60\%} \\
    CIFAR10-Torjan & 81.81\% & 99.50\% & 74.72\% \small{($\Downarrow$8.7\%)} & 6.50\% & 27.15\% \small{($\Downarrow$66.8\%)} & 3.50\% & 35.5\% \small{($\Downarrow$56.6\%)} & 21.80\% & \boldmath{}\textbf{81.81\% \small{($\Downarrow$0.0\%)}}\unboldmath{} & \textbf{0.20\%} \\
    CIFAR10-Hidden & 82.67\% & 99.40\% & 74.31\% \small{($\Downarrow$10.1\%)} & 9.20\% & 37.27\% \small{($\Downarrow$55.0\%)} & 3.73\% & 51.64\% \small{($\Downarrow$37.5\%)} & 0.80\% & \boldmath{}\textbf{82.67\% \small{($\Downarrow$0.0\%)}}\unboldmath{} & \textbf{0.30\%} \\
    \midrule
    Average & 88.92\% & 98.83\% & 75.41\% \small{($\Downarrow$15.2\%)} & 8.28\% & 68.58\% \small{($\Downarrow$22.9\%)} & 7.62\% & 73.14\% \small{($\Downarrow$16.6\%)} & 9.50\% & \boldmath{}\textbf{88.85\% \small{($\Downarrow$0.1\%)}}\unboldmath{} & \textbf{0.46\%} \\
    \bottomrule
    \end{tabular}%
    }
  \label{tab:backdoor}%
\end{table*}%

\begin{table*}[tbp]
  \centering
  \renewcommand{\arraystretch}{0.9}
  \setlength{\abovecaptionskip}{2pt} 
  \setlength{\belowcaptionskip}{2pt}
  \caption{Model accuracy (Acc, the higher, the better) and unfairness (UF, the lower, the better) on independent test sets before and after applying different debugging methods.}
  \setlength{\tabcolsep}{1.5mm}{
    \begin{tabular}{c|cc|cc|cc|cc|cc}
    \toprule
    \multirow{2}[4]{*}{\textbf{Model}} & \multicolumn{2}{c|}{\textbf{Original}} & \multicolumn{2}{c|}{\textbf{ADF}} & \multicolumn{2}{c|}{\textbf{CARE}} & \multicolumn{2}{c|}{\textbf{IDNN}} & \multicolumn{2}{c}{\textbf{\tool}} \\
\cmidrule{2-11}          & Acc   & UF    & Acc   & UF    & Acc   & UF    & Acc   & UF    & Acc   & UF \\
    \midrule
    \midrule
    Census-race & 84.92\% & 7.00\% & 83.08\% \small{($\Downarrow$2.2\%)} & 1.03\% & 81.57\% \small{($\Downarrow$3.9\%)} & 0.81\% & 81.27\% \small{($\Downarrow$4.3\%)} & \textbf{0.07\%} & \boldmath{}\textbf{84.86\% \small{($\Downarrow$0.1\%)}}\unboldmath{} & 0.09\% \\
    Census-gender & 84.92\% & 4.53\% & 82.31\% \small{($\Downarrow$3.1\%)} & 0.92\% & 82.89\% \small{($\Downarrow$2.4\%)} & 0.17\% & 81.82\% \small{($\Downarrow$3.7\%)} & 0.21\% & \boldmath{}\textbf{84.92\% \small{($\Downarrow$0.0\%)}}\unboldmath{} & \textbf{0.00\%} \\
    Census-age & 84.92\% & 19.04\% & 78.77\% \small{($\Downarrow$7.2\%)} & 2.31\% & 81.27\% \small{($\Downarrow$4.3\%)} & 4.90\% & 83.77\% \small{($\Downarrow$1.4\%)} & 9.77\% & \boldmath{}\textbf{84.74\% \small{($\Downarrow$0.2\%)}}\unboldmath{} & \textbf{0.11\%} \\
    \midrule
    COMPAS-gender & 68.36\% & 12.49\% & 61.56\% \small{($\Downarrow$9.9\%)} & 6.21\% & 67.12\% \small{($\Downarrow$1.8\%)} & 2.68\% & 66.79\% \small{($\Downarrow$2.3\%)} & 1.15\% & \boldmath{}\textbf{68.25\% \small{($\Downarrow$0.2\%)}}\unboldmath{} & \textbf{0.00\%} \\
    COMPAS-race & 68.36\% & 8.47\% & 66.79\% \small{($\Downarrow$2.3\%)} & 5.77\% & 67.65\% \small{($\Downarrow$1.0\%)} & 0.44\% & 67.66\% \small{($\Downarrow$1.0\%)} & 0.00\% & \boldmath{}\textbf{68.74\% \small{($\Uparrow$0.6\%)}}\unboldmath{} & \textbf{0.00\%} \\
    \midrule
    Bank-age & 81.46\% & 12.30\% & 71.25\% \small{($\Downarrow$12.5\%)} & 3.03\% & 80.68\% \small{($\Downarrow$1.0\%)} & 4.64\% & 81.37\% \small{($\Downarrow$0.1\%)} & 6.97\% & \boldmath{}\textbf{81.37\% \small{($\Downarrow$0.1\%)}}\unboldmath{} & \textbf{0.10\%} \\
    Bank-edu & 81.46\% & 5.71\% & 73.35\% \small{($\Downarrow$10.0\%)} & 0.56\% & 80.37\% \small{($\Downarrow$1.3\%)} & 0.57\% & 81.28\% \small{($\Downarrow$0.2\%)} & 0.81\% & \boldmath{}\textbf{81.28\% \small{($\Downarrow$0.2\%)}}\unboldmath{} & \textbf{0.00\%} \\
    \midrule
    Average & 79.20\% & 9.93\% & 73.87\% \small{($\Downarrow$6.7\%)} & 2.83\% & 77.36\% \small{($\Downarrow$2.3\%)} & 2.03\% & 77.71\% \small{($\Downarrow$1.9\%)} & 2.71\% & \boldmath{}\textbf{79.17\% \small{($\Downarrow$0.0\%)}}\unboldmath{} & \textbf{0.04\%} \\
    \bottomrule
    \end{tabular}%
    }
  \label{tab:unfairness}%
\end{table*}%

\subsubsection{RQ1: Effectiveness and Efficiency}

\textbf{\\1) Effectiveness in mitigating backdoor attacks.}  We evaluate \tool’s capability to mitigate backdoor attacks while preserving model functionality on 3 benchmarks under three 3 attack strategies. Model accuracy (Acc) is measured on the original clean test set, while the backdoor attack success rate (ASR)~\cite{gu2019badnets}, which is defined as the percentage of triggered inputs misclassified to the attacker-chosen target class, is evaluated on a separate subset of the test data attacked by the above strategies. Both sets are independent of the debugging and training data to ensure unbiased evaluation.

\textit{Results.}  As shown in Table~\ref{tab:backdoor}, \tool achieve near-perfect backdoor mitigation with an average ASR of 0.46\% while maintaining the original model accuracy with an average accuracy drop of 0.1\%, significantly outperforming all baselines. For instance, on CIFAR10-BadNets, \tool reduces ASR to 0.6\% without sacrificing accuracy, even slightly improving it by 0.2\%, whereas IDNN and CARE degrade accuracy by 35.0\% and 33.3\%, respectively. Notably, \tool completely eliminates BadNets attacks on MNIST with an ASR of 0.00\% while retaining 100\% original accuracy. In contrast, Fine-Pruning severely compromises model functionality with an average accuracy drop of 15.2\%, especially on MNIST-Trojan where accuracy plummets by 35.8\%.

\textit{Analysis.} The superiority of \tool comes from its targeted intervention in the attention map, which serve as the primary carrier of backdoor logic. In contrast, the three baselines fail to effectively address the spatially distributed backdoor patterns encoded in attention maps. The poor performance of IDNN on the CIFAR-10 indicates that isolating neurons alone is insufficient to eliminate triggers distributed across multiple attention maps. Additionally, the instability of Fine-Pruning and CARE under different experimental configurations further supports existing research findings~\cite{clark2019does,vig2019analyzing} that the attention mechanism causes neurons to exhibit different behaviors under varying inputs. Consequently, editing a subset of neurons to handle different inputs becomes an unstable approach. Notably, even when IDNN isolates 80\% of neurons under specific configurations, the remaining attention weights can still propagate trigger-related features, leading to a high ASR on CIFAR10. In contrast, \tool successfully overcomes these issues by redistributing attention map at a more fundamental level.

Furthermore, the significant accuracy drop in baseline methods reveals their overly aggressive repair strategies. Fine-Pruning and IDNN prune or isolate neurons, while CARE optimizes neuron parameters, all of them risk inadvertently altering critical original feature representations. As a result, these methods uniformly apply modified parameters to all inputs during inference, inevitably contaminating the features of clean data and degrading the model's original performance. \tool mitigates this issue through its \textit{Detector} selection mechanism, redistributing attention maps only when anomalies are reported.

\noindent\textbf{2) Effectiveness in mitigating unfairness.} We evaluate \tool’s capability to mitigate model unfairness (UF) across 3 commonly used fairness-critical datasets. We define UF as the proportion of test instances where predictions change when only protected attributes are modified while non-protected features remain constant. It is computed by perturbing protected attributes (Eq.~\ref{eq:uf}) and checking for prediction inconsistencies rate. Both evaluations use data independent of the debugging and training sets to ensure the model and \tool have no prior exposure.

\textit{Results.} As shown in Table~\ref{tab:unfairness}, \tool reduces unfairness to near-zero levels, achieving an average UF of 0.04\% while maintaining negligible accuracy loss of 0.0\% on average, surpassing all baselines. For example, on Census-age, \tool achieves 0.11\% UF compared to the original 19.04\% with only 0.2\% accuracy degradation, whereas ADF and CARE degrade accuracy by 7.2\% and 4.3\%, respectively. \tool fully eliminates unfairness in four scenarios such as COMPAS-race and Bank-edu while preserving original accuracy. Notably, IDNN struggles to balance fairness and utility. Though it achieves 0.07\% UF on Census-race, its accuracy drops 4.3\%, far exceeding \tool’s 0.1\% drop.

\textit{Analysis.} \tool’s superiority stems from its ability to identify and suppress attention overflows toward protected attributes, which often serve as latent decision shortcuts. For instance, in COMPAS-gender, the original model’s high UF of 12.49\% implies it disproportionately attends to gender-related tokens during inference. 
Baselines like ADF attempts to reduce bias by retraining on augmented discriminatory instances, but this disrupts feature representations, causing a 9.9\% accuracy drop on COMPAS-gender.
In contrast, \tool surgically neutralizes biased attention columns without altering the model’s parameters. 

\noindent\textbf{3) Efficiency of \tool.} To assess the efficiency of \tool compared to the baseline methods, we recorded the execution times of key components for each method across all experimental configurations and computed their average values.

\textit{Results.} The average time costs reveal stark differences. IDNN requires 21 minutes on average excluding T2T-ViT on CIFAR10 to isolate critical neurons, primarily due to its each-neuron contribution analysis. Specifically, for large models like T2T-ViT on CIFAR10 
, this process exceeds 3 hours.
CARE spends more than 1 hour in total due to its iterative weight adjustment algorithm.
Fine-Pruning takes 2 minutes for pruning but necessitates 43 additional minutes for retraining to recover accuracy.
ADF requires 18 minutes for discriminatory sample generation and 11 minutes for fairness-aware fine-tuning.

To assess the efficiency of \tool compared to baseline methods, we measured the total time cost of its key components, including both the offline preparation and online inference phases. Unlike baselines that often require extensive retraining or parameter editing, \tool performs lightweight attention-level patching. We provide details of the offline cost. The total offline time to prepare the \textit{Detector} consists of three parts: \textbf{1) Debugging Set Construction:} This includes identifying clean-backdoor or clean-bias pairs using existing techniques, which takes approximately 13s per configuration. \textbf{2) Interactions with Transformers to collect the Attention maps:} This step involves forwarding selected samples through the original model to extract attention maps from all Transformer layers. This process is streamlined and takes less than 30s on average. \textbf{3) Detector Training:} The training procedure, based on contrastive learning, is lightweight and converges within 1 minute on average. Overall, the total offline cost is significantly lower than most baselines. Moreover, this offline phase is a one-time cost and decoupled from the model, making \tool easy to reuse it across similar settings.

During the online debugging process, if \tool's \textit{Detector} reports anomalies, detecting and redistributing over-attention introduces an average delay of 0.8 ms per sample. If no anomaly is reported, the average delay per sample is 0.1 ms. In comparison, the original inference time is 0.5 ms per sample.

\textit{Analysis.} The efficiency of \tool in real-world applications stems from its design. First, its offline phase avoids the computationally expensive neuron analysis in IDNN and the iterative optimization in CARE. Second, performing online debugging on attention maps reduces computational costs, whereas IDNN and CARE require extensive neuron computations, and Fine-Pruning and ADF necessitate retraining. Moreover, we consider that an additional 0.8 ms delay for samples exhibiting backdoor attacks or fairness violations is acceptable in safety-critical environments. Since clean inputs that only require detection constitute the vast majority of real-world workloads, the overall inference overhead remains minimal in practice.

\begin{center}
\fcolorbox{black}{gray!10}{\parbox{.95\linewidth}{\textbf{Answer to RQ1:}
  \tool outperforms baseline methods in debugging transformer-based models while preserving their original functionality and demonstrating superior efficiency.}
  }
\end{center}

\subsubsection{RQ2: Performance of the Detector}

\label{sec:exp-det}

\textbf{\\} To rigorously evaluate the capability of \tool’s \textit{Detector}, we construct an evaluation set containing both backdoor/fairness-violating data and clean data. Specifically, we use the backdoor/fairness-violating data from the test sets in RQ1 (independent of the debugging and training sets) and pair them with an equal number of clean test data to ensure balanced positive/negative classes. For the output of the \textit{Detector}, the evaluation is performed at the attention map level, with strict criteria: (1) For each over-attention map, a true positive (TP) is counted only if the \textit{Detector} identifies all ground-truth anomalous columns without any false alarms. Any deviation (e.g., missing an anomalous column or flagging a normal column) is considered a false negative (FN). (2) For each clean attention map, a true negative (TN) is counted only if the \textit{Detector} reports nothing; otherwise, it is a false positive (FP). Due to space constraints, we report results on 6 challenging configurations from RQ1 where \tool exhibited slightly degraded performance.

\begin{table}[tbp]
  \centering
  \setlength{\abovecaptionskip}{2pt} 
  \setlength{\belowcaptionskip}{2pt}
  \caption{Performance metrics of the \tool's \textit{Detector}.}
    \setlength{\tabcolsep}{2.3mm}{\begin{tabular}{c|ccc|cc}
    \toprule
    \textbf{Model} & \textbf{P} & \textbf{R} & \textbf{F1} & \textbf{FPR} & \textbf{FNR} \\
    \midrule
    MNIST-Hidden & 0.981 & 0.994 & 0.987 & 0.020 & 0.006 \\
    FASHION-Torjan & 0.945 & 0.999 & 0.971 & 0.059 & 0.001 \\
    CIFAR10-BadNets & 0.924 & 1.000 & 0.960 & 0.082 & 0.000 \\
    Census-age & 0.962 & 0.995 & 0.978 & 0.040 & 0.005 \\
    COMPAS-race & 0.910 & 1.000 & 0.953 & 0.099 & 0.000 \\
    Bank-edu & 0.952 & 1.000 & 0.975 & 0.051 & 0.000 \\
    \bottomrule
    \end{tabular}%
    }
  \label{tab:detector}%
  \vspace{-8pt}
\end{table}%

\begin{figure}[tbp]
  \setlength{\abovecaptionskip}{2pt} 
  \setlength{\belowcaptionskip}{2pt}
    \centering
    \subfloat{%
        \includegraphics[width=0.24\textwidth]{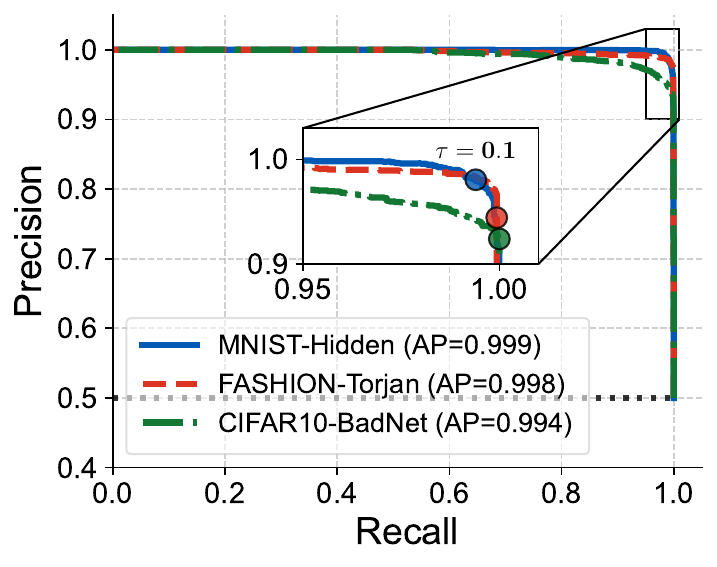}%
        \label{fig:image1}%
    }
    \subfloat{%
        \includegraphics[width=0.24\textwidth]{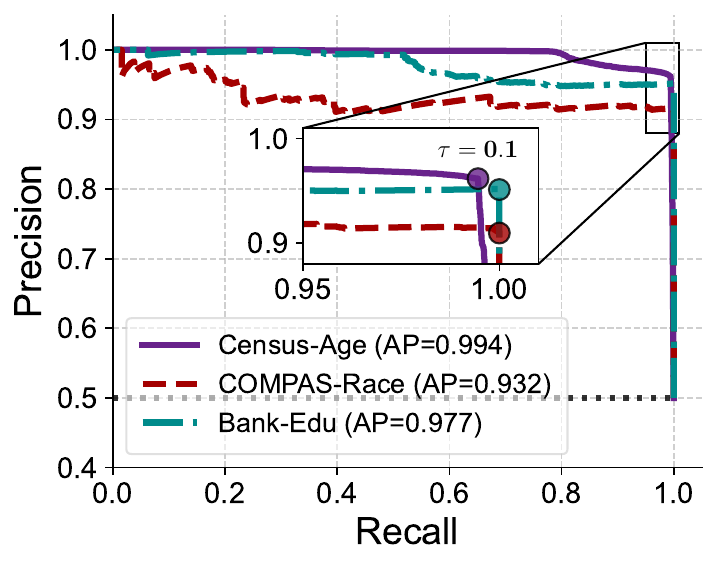}%
        \label{fig:image2}%
    }
    \caption{PR-curve of \tool's \textit{Detector} performance.}
    \label{fig:prcurve}
    \vspace{-8pt}
\end{figure}

\textit{Results.} As shown in Table~\ref{tab:detector}, the \textit{Detector} achieves near-perfect precision between 0.910 and 0.981, with recall consistently exceeding 0.994 across all evaluated scenarios. Its F1-scores remain above 0.95 throughout. Notably, the false positive rate stays below 0.1, varying from 0.020 to 0.099, while the false negative rate is negligible, never exceeding 0.006. Figure~\ref{fig:prcurve} further validates the robustness through precision-recall curves, which cluster tightly in the upper-right corner, indicating high precision across different recall levels. The operating point at threshold $\tau = 0.1$ lies close to the balanced region~\cite{davis2006relationship}, demonstrating stable performance under \tool’s default configuration.

\textit{Analysis.} These results conclusively answer RQ2.
First, the near-perfect recall and negligible false negative rate indicate that the \textit{Detector} rarely misses true anomalous columns, even in challenging scenarios like CIFAR10-BadNet, where triggers are embedded in visually complex backgrounds. This robustness stems from the contrastive training paradigm: by explicitly contrasting normal and anomalous attention maps in the \textit{Debugging Set}, the \textit{Detector} learns to isolate discriminative features such as localized spikes in attention weights associated with backdoor or biased behavior.
Second, the low false positive rate, which is at most 0.099, ensures that clean attention maps are largely unperturbed. For example, in MNIST-Hidden, where triggers are semi-transparent and visually subtle, the \textit{Detector} achieves a false positive rate of only 0.02, demonstrating its ability to distinguish between benign feature selection and malicious over-attention. This specificity is critical for preserving model functionality, as excessive false positives could degrade accuracy on clean inputs, a common limitation in neuron-editing techniques.
This dual capability of high sensitivity to backdoor or fairness-violating patterns and strong specificity to benign patterns validates the \textit{Detector}’s design rationale: leveraging contrastive learning on \textit{Debugging Set} enables discriminative feature extraction for attention anomaly localization.

\begin{center}
\fcolorbox{black}{gray!10}{\parbox{.95\linewidth}{\textbf{Answer to RQ2:}
  \tool's \textit{Detector} can accurately identify over-attention while maintaining a low false positive rate on clean data, thereby preserving the model's functionality.}
  }
\end{center}

\begin{figure*}[tbp]
  \setlength{\abovecaptionskip}{2pt} 
  \setlength{\belowcaptionskip}{2pt}
    \centering
    \subfloat{%
        \includegraphics[width=0.165\textwidth]{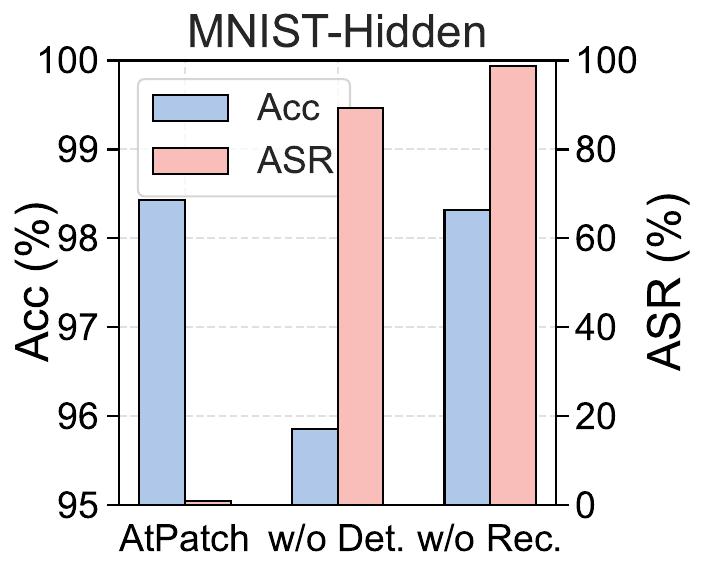}%
        \label{fig:image1}%
    }
    \hfill
    \subfloat{%
        \includegraphics[width=0.165\textwidth]{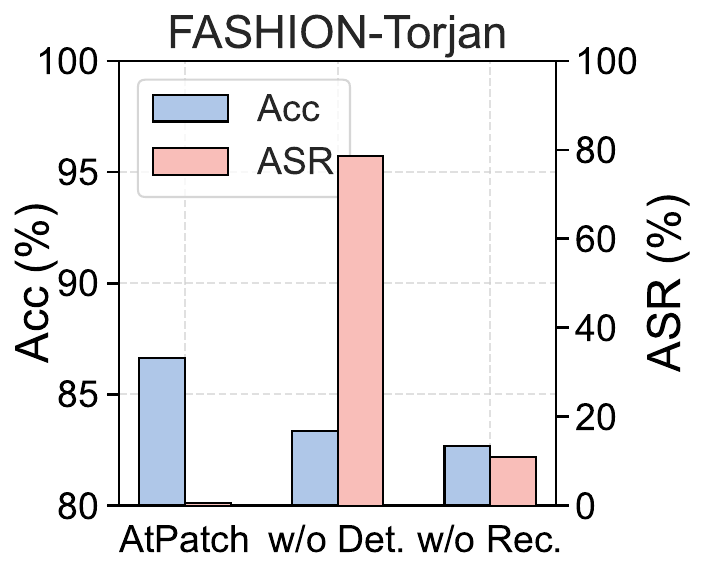}%
        \label{fig:image2}%
    }
    \hfill
    \subfloat{%
        \includegraphics[width=0.165\textwidth]{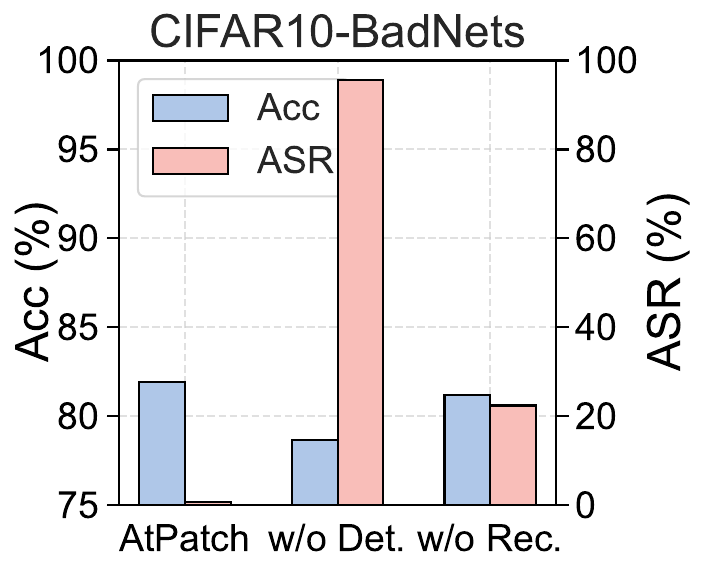}%
        \label{fig:image3}%
    }
    \hfill
    \subfloat{%
        \includegraphics[width=0.165\textwidth]{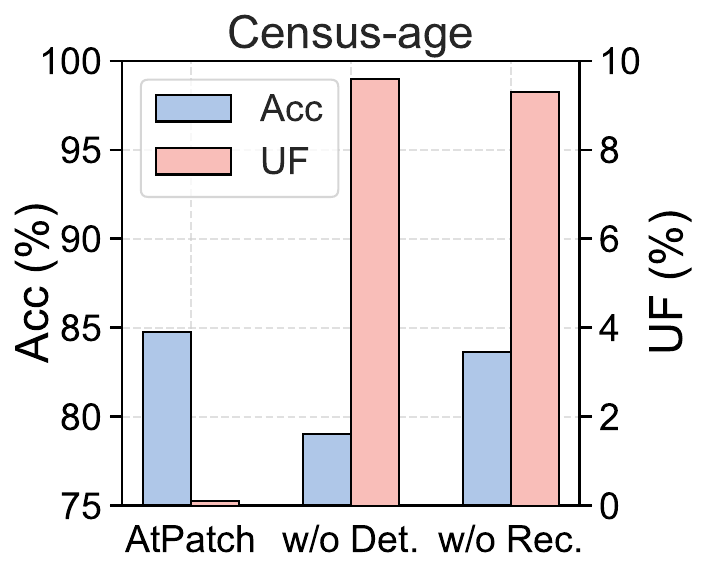}%
        \label{fig:image4}%
    }
    \hfill
    \subfloat{%
        \includegraphics[width=0.165\textwidth]{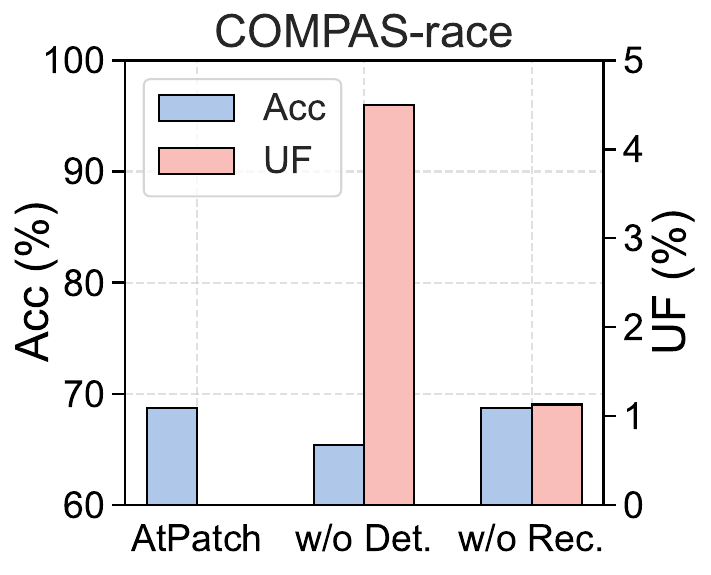}%
        \label{fig:image5}%
    }
    \hfill
    \subfloat{%
        \includegraphics[width=0.165\textwidth]{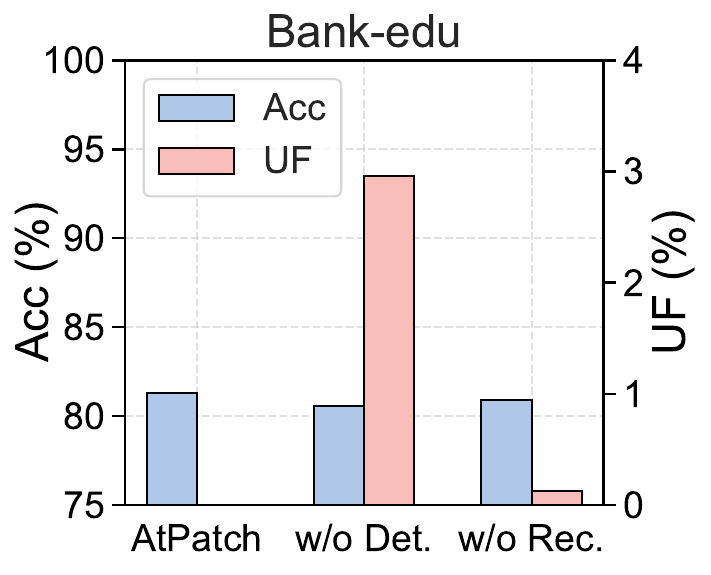}%
        \label{fig:image6}%
    }
    \caption{Results of Acc, ASR, and UF on \tool and its two variants.}
    \label{fig:ab}
    \vspace{-8pt}
\end{figure*}

\subsubsection{RQ3: Ablation Study}

\textbf{\\} To evaluate the contribution of key components in \tool, we conducted an ablation study by systematically removing key elements of the framework. Specifically, we identified two critical components: the \textit{Detector}, responsible for identifying over-attention columns in the attention map, and the \textit{rescaling process}, which rescales the unaffected columns to redistribute the attention. We design two variants: \textbf{w/o Det.} replaces the trained \textit{Detector} with a random selector that picks 1–3 columns (matching the average number of anomalies reported by \tool) for each attention map, and \textbf{w/o Rec.} directly overwrites detected anomalous columns without rescaling other values. Due to space constraints, we report results on the datasets and models used in RQ2.

\textit{Results.} As shown in Figure~\ref{fig:ab}, the ablation results reveal significant performance degradation in both variants.
For backdoor attacks, \textbf{w/o Det.} exhibits high ASR, reaching 89.3\% on MNIST-Hidden compared to \tool’s mere 0.9\%, and also suffers from reduced accuracy, dropping to 78.65\% on CIFAR10-BadNets versus \tool’s 81.92\%. These results indicate that random column replacement fails to mitigate backdoor attacks effectively. Similarly, \textbf{w/o Rec.} remains largely ineffective, with ASR as high as 22.37\% on CIFAR10-BadNets, whereas \tool reduces it to just 0.6\%, suggesting that improper attention redistribution leaves residual over-attention. 
Besides, for unfairness mitigation, \textbf{w/o Det.} leads to substantial unfairness, reaching 9.6\% on Census-age compared to \tool’s minimal 0.11\%. Meanwhile, \textbf{w/o Rec.} still retains notable bias, with 9.31\% unfairness on Census-age.
Notably, even when \textbf{w/o Rec.} occasionally achieves low unfairness, such as 1.13\% on COMPAS-race, its accuracy drops by 0.6\% compared to \tool, suggesting unstable functional preservation.

\textit{Analysis.} The results confirm that both components are indispensable. The \textit{Detector}’s precision is critical: random column adjustments in \textbf{w/o Det.} degrade accuracy without resolving attacks or bias, proving that targeted anomaly localization, not arbitrary modifications, drives \tool’s efficacy. Meanwhile, the rescaling process can suppress the over-attention of anomalous columns. Without this process, directly replacing anomalous columns would cause the unaffected columns to retain their original low attention. This results in the replaced columns still exhibiting over-attention relative to others, leading to high ASR or UF. Moreover, such an attention map would lead to feature aggregation distortion, thereby compromising the model's original functionality and resulting in low accuracy. 
This highlights the critical importance of \tool's two-stage design (detection and redistribution) for reliable and uninterrupted hot-fix.

\begin{center}
\fcolorbox{black}{gray!10}{\parbox{.95\linewidth}{\textbf{Answer to RQ3:}
  Both the \textit{Detector} and the \textit{rescaling process} are indispensable, as their absence severely degrades \tool’s debugging efficacy and functional preservation.}
  }
\end{center}

\vspace{-3pt}
\section{Discussion}
\vspace{-3pt}
\subsection{Orthogonality with Debugging Neurons}
While existing neuron-editing debugging techniques~\cite{chen2024isolation,sun2022causality,liu2018fine} primarily focus on identifying and modifying defective neurons or parameters, our \tool operates at a distinct level by redistributing intermediate attention maps during inference. This fundamental difference in intervention targets establishes the orthogonality between \tool and neuron-based debugging methods. 

However, previous studies~\cite{wu2021adversarial,liu2018fine} have shown that neuron editing methods often require sacrificing part of the model's functionality to mitigate defects, as parameter modifications inevitably disrupt the original decision logic. In contrast, \tool is designed to better preserve the model's original functionality. The \textit{Detector} in \tool selectively intervenes in the attention map, ensuring more targeted modifications. Additionally, existing study~\cite{dai2021knowledge} also suggests that neuron editing techniques are not well-suited for complex transformer architectures. This is because the dynamic computation graphs constructed by the attention mechanism cause neurons to exhibit different behaviors under different inputs, making fixed neuron edits potentially ineffective across all inputs~\cite{clark2019does,vig2019analyzing}. In contrast, \tool’s hot-fix capability enables real-time redistribution of attention directly at the more fundamental attention map level. Our experimental results further demonstrate this. Using only \tool can achieve superior mitigation of backdoor attacks and unfairness while maintaining high model accuracy, outperforming existing methods. Moreover, the results highlight the instability of existing approaches when applied to transformer architectures. Given these findings, along with the computational overhead introduced by combining multiple debugging frameworks, this paper focuses on evaluating \tool as an independent solution.

Nevertheless, interventions at the attention map level and the neuron level are architecturally separate and do not present any fundamental conflicts. Therefore, we encourage future research to propose new, more effective and efficient neuron editing methods tailored for transformer architectures. Such advancements would make the hybridization of these two approaches highly valuable in multi-level defect mitigation scenarios.

\vspace{-3pt}
\subsection{OOD Robustness of \tool}
Out-of-distribution (OOD) data presents a general and well-known challenge for learning-based approaches. In the context of \tool, we identify two following distinct OOD scenarios: 

\textbf{1) OOD in input distribution (e.g., new domains of images or text).} Since \tool operates on attention maps rather than raw inputs, it is inherently more robust to shifts in input distribution. As long as the model exhibits over-attention patterns, the \textit{Detector} can still effectively localize anomalies. In our additional experiments, a \textit{Detector} trained on MNIST was applied to FASHION (simulating cross-domain scenarios) and still achieve a low attack success rate of 1.13\%, close to the original 0.90\%. This suggests that \tool generalizes well under moderate input distribution shifts.

\textbf{2) OOD in over-attention patterns (e.g., new backdoor triggers or protected attribute)}. This type of OOD reflects the victim model itself has been changed (e.g., poisoned by new triggers). In such cases, the previous \textit{Debugging Set} of \textit{Detector} may no longer reflect the model's current behavior. To address this, developers should update the \textit{Debugging Set} using the inverse optimization method described in Section~\ref{sec:debuggingset} and retrain the \textit{Detector} accordingly. This process remains lightweight and does not require changing the model’s parameters.

\textbf{Alternative to retraining.} As a potential direction for future work, we consider replacing the supervised \textit{Detector} with unsupervised anomaly detection techniques that can identify over-attention without requiring a labeled \textit{Debugging Set}. This could eliminate the need for retraining when the model evolves. However, such methods are known to suffer from high false positive rates, which may lead to unnecessary or incorrect interventions. Balancing generalization and precision remains an open challenge in this area.

\vspace{-6pt}
\subsection{The Scalability of \tool}

\tool is a general-purpose debugging framework designed for transformer-based models. Its core mechanism operates directly on attention maps, making it inherently applicable across different data modalities such as images, tabular data, and natural language.

Our main evaluation focused on image and tabular datasets to ensure consistency with prior works~\cite{chen2024isolation,liu2018fine,sun2022causality,zhang2020white,monjezi2023information,dasu2024neufair} and enable fair comparisons. However, \tool is not limited to these domains. To demonstrate its extensibility, we applied \tool to a BERT-based sentiment classifier on the IMDB dataset under the BadNL~\cite{chen2021badnl} backdoor attack. \tool reduced the attack success rate from 97.90\% to 0.70\%, while maintaining clean accuracy (from 87.63\% to 87.34\%). These findings support \tool’s ability to scale to larger models and more complex modalities. We believe this provides a strong foundation for future work on applying \tool to multi-modal transformers and other attention-related vulnerabilities in real-world systems.

\vspace{-6pt}
\subsection{Threats to Validity}
External threats arise from potential biases in datasets, model architectures, backdoor attack strategies and the selection of protected attributes. To mitigate this, we evaluate \tool on 6 widely recognized benchmarks commonly used in prior debugging studies and 6 different model architectures, 3 prevalent backdoor attack paradigms, and widely adopted 4 fairness-sensitive attributes. Internal threats relate to potential implementation errors in \tool or baselines. To mitigate this, we rigorously validate \tool’s implementation and reuse official codebases for all baselines as much as possible.
Construct threats relate to hyperparameter setting, particularly the $\lambda$ of contrastive loss and detection threshold $\tau$. To mitigate this, we align contrastive learning hyperparameters with established best practices. The threshold $\tau$ is empirically determined, and its robustness is validated in RQ2.

Adaptive Attack Resilience. Another potential threat involves adaptive attackers which may attempt to craft triggers that evade \tool’s detection and attention redistribution. Unlike traditional adversarial attacks that directly perturb the input, attackers in our setting must indirectly influence the attention map through input manipulation, which is significantly more complex, especially in black-box scenarios where the attention behavior is not observable. To further enhance robustness, one promising direction is to employ an ensemble of \textit{Detector} with diverse architectures or training objectives. This would increase the difficulty for an attacker to simultaneously bypass all detection mechanisms. We acknowledge this as an important avenue for future work.

\vspace{-6pt}
\section{Related Works}

\textit{DNN Testing.} Research on DNN testing has explored various coverage criteria to assess model robustness~\cite{feng2020deepgini,weng2023prioritizing,sun2018testing}. DeepXplore~\cite{pei2017deepxplore} and DeepGauge~\cite{ma2018deepgauge} introduced neuron activation coverage metrics to guide test case generation. Subsequent work like NeuronInspect~\cite{huang2019neuroninspect} extended these ideas to backdoor detection by analyzing activation patterns. Furthermore, DeepTest~\cite{tian2018deeptest} specialized in autonomous driving scenarios, generating diverse synthetic inputs to expose erroneous behaviors. DeepMutation~\cite{ma2018deepmutation} formalized mutation testing for DNNs, quantifying test suite quality by measuring fault detection rates. DeepConcolic~\cite{sun2019deepconcolic} combined symbolic execution and concolic testing to maximize decision boundary coverage. While these methods excel at defect identification, they primarily target pre-deployment testing and lack runtime repair mechanisms. In contrast, \tool extends testing principles to runtime monitoring and intervention, addressing defects dynamically without disrupting model deployment.

\noindent\textit{DNN Debugging.} Debugging techniques for DNNs aim to diagnose and mitigate defects such as backdoors~\cite{liu2019abs,gao2019strip} or unfairness~\cite{tramer2017fairtest,angell2018themis}. IDNN~\cite{chen2024isolation} isolates critical neurons responsible for defects through dynamic contribution analysis and repairs them without retraining, while CARE~\cite{sun2022causality} combines causal attribution with particle swarm optimization to adjust neuron weights for bias elimination. For backdoor mitigation, methods like Fine-Pruning~\cite{liu2018fine} prune suspicious neurons and fine-tune models, and Neural Cleanse~\cite{wang2019neural} detects triggers via activation anomalies. ADF~\cite{zhang2020white} addresses unfairness by generating discriminatory instances via adversarial sampling and retraining. However, these methods often require architectural modifications, weight adjustments, or trigger reconstruction, which may compromise model functionality or demand retraining. \tool distinguishes itself by operating non-invasively: it identifies over-attention patterns via contrastive learning, dynamically \textit{patches} attention maps at runtime, and preserves the original model’s parameters and behavior on clean inputs.

\vspace{-6pt}
\section{Conclusion}
In this paper, we propose \tool, a novel hot-fix method for transformer-based DNNs, inspired by over-attention phenomenon, and software engineering paradigms of delta debugging and hot patching. \tool detects and redistributes anomalous attention distributions at inference time, mitigating backdoor attacks and unfairness without modifying model parameters. Specifically, \tool employs a \textit{Detector} to identify over-attention columns. It then replaces the detected anomalous columns with unified benign attention columns and rescales the unaffected columns to redistribute attention. Compared to traditional methods, \tool is more specifically tailored for transformer-based models. Its on-the-fly design eliminates the need to modify model parameters or retrain the model, making it more suitable for deployed models. Experimental results demonstrate that \tool can effectively debug the model while preserving its original functionality.

\vspace{-5pt}
\begin{acks}
We would like to thank anonymous reviewers for their construc-
tive comments. This project was partially funded by the National Key R\&D Program of China (No. 2024YFB4506400), the National Research Foundation, Singapore, and the Cyber Security Agency under its National Cybersecurity R\&D Programme (NCRP25-P04-TAICeN), and the National Natural Science Foundation of China under Grant No. 62372225 and No. 62272220. Any opinions, findings and conclusions or recommendations expressed in this material are those of the author(s) and do not reflect the views of National Research Foundation, Singapore and Cyber Security Agency of Singapore.
\end{acks}

\bibliographystyle{ACM-Reference-Format}
\bibliography{sample-base}

\end{document}